\documentstyle[preprint,aps,prl]{revtex}
\draft

\begin{document}
\title{Condensation in Globally Coupled Populations \\
of Chaotic Dynamical Systems}
\author{D.H. Zanette \thanks{Permanent address: Consejo Nacional
de Investigaciones Cient\'{\i}ficas y T\'ecnicas, Centro At\'omico Bariloche,
8400 Bariloche, Argentina} and A. S. Mikhailov}
\address{Fritz-Haber-Institut der Max-Planck-Gesellschaft\\
Faradayweg 4-6, 14195 Berlin, Germany}
\date{\today}
\maketitle

\begin{abstract}
The condensation transition, leading to complete mutual synchronization in
large populations of globally coupled chaotic R\"{o}ssler oscillators, is
investigated. Statistical properties of this transition and the cluster
structure of partially condensed states are analyzed.
\end{abstract}


\newpage

Appearance of condensates is an important property of quantum and classical
systems. When a population of interacting Bose particles (such as, e.g.,
superfluid liquid He$^4$) is considered, a condensate is formed by a
finite fraction of particles occupying the same quantum state \cite{Bose}.
Condensation of electron-hole pairs in metals is responsible for the
phenomenon of superconductivity \cite{superc}. Condensation phenomena in
classical systems are closely related to the effect of mutual
synchronization of oscillations. In large populations of globally coupled
limit-cycle oscillators, one or several groups of oscillators following
exactly the same phase trajectory can spontaneously appear 
\cite{Kura,Naka,Saka,Hakim,Golomb}. 
A similar behavior in populations of globally coupled logistic or circle maps 
has been found by K. Kaneko \cite{Kaneko1,Kaneko2}. 
Mutual synchronization of several coupled chaotic oscillators has been
studied (see, e.g., \cite{Pec}). To analyze properties of the
condensation transition, which involves progressive formation of coherent
oscillator groups, larger systems must be however considered. In this Letter
we investigate   condensation phenomena in a population of a thousand
globally coupled chaotic R\"{o}ssler oscillators.

Not every form of global coupling between  elements can lead to mutual
synchronization. Synchronous chaotic oscillations are possible if  
`vector' coupling (with identical coupling coefficients for all oscillator
variables) is employed \cite{fuji}. Asymptotic synchronization conditions in
the limit of high coupling intensities have also been constructed for the
case when coupling involves only one of the oscillator variables \cite{Pec}.
However, only the linear stability of synchronous oscillations is proven in
both of these cases and therefore even for a strong coupling the
presence of other dynamical attractors, corresponding to persistent
asynchronous oscillations, cannot be excluded \cite{Chate}.

In this Letter a different form of global coupling that yields more robust
synchronization is introduced. We take an arbitrary dynamical system with $m$
variables ${\bf w}(t)=\{{\bf w}_1(t),...,{\bf w}_m(t)\}$ whose evolution
is determined by equations 
\begin{equation}
\dot{\bf w}={\bf f}({\bf w}),  \label{7}
\end{equation}
where ${\bf f}({\bf w})$ are some known functions. A globally coupled
population consisting of $N$ such dynamical systems ($i=1,2,...,N)$ is
constructed by adding the coupling terms, so that the resulting evolution
equations are 
\begin{equation}
\dot{{\bf w}}_i={\bf f}({\bf w}_i)+\varepsilon \hat{{\bf A}}(\bar{\bf w}
-{\bf w}_i)+\varepsilon'[{\bf f}(\bar{\bf w})-{\bf f}
({\bf w}_i)],  \label{8}
\end{equation}
where $\bar{\bf w}$ is the global average, $\bar{\bf w}(t)=N^{-1}\sum_{i=1}^N
{\bf w}_i(t)$, the positive parameters $\varepsilon$ and $\varepsilon'$ specify 
the intensity of global coupling, and $\hat{\bf A}$ is a constant matrix. Note 
that the previously used forms of coupling correspond to a choice $\varepsilon'
=0$; the `vector' coupling is obtained for $\hat{\bf A}={\bf I}$ where 
${\bf I}$ is the identity matrix.

When $\varepsilon'=1$ equations (\ref{8}) reduce to 
\begin{equation}
\dot{\bf w}_i=\varepsilon \hat{{\bf A}}(\bar{\bf w}-{\bf w}_i)+{\bf f}
(\bar{\bf w}).
\end{equation}
They describe linear motion of elements forming the population under action
of a force which is collectively produced by all of them. The evolution of
the deviations $\delta {\bf w}_i(t)={\bf w}_i(t)-\bar{\bf w}(t)$ is
then determined by an {\it exact} linear set of equations $\dot{\delta 
{\bf w}_i}=-\varepsilon \hat{{\bf A}}\delta {\bf w}_i$. Therefore, these
deviations exponentially decrease with time, and global stability of
synchronous oscillations is thus guaranteed, if all real eigenvalues of the
matrix $\hat{\bf A}$ and the real parts of all its complex eigenvalues are
positive.

Since globally stable synchronous oscilations are present in the system 
(\ref{8}) at $\varepsilon'=1$, they may also be expected in a
neighborhood of this point, i.e. for an interval of coupling intensities 
$\varepsilon_0<\varepsilon'<1$. As shown below, this interval
can be so wide that even weak coupling leads to the synchronization.

We consider a population of chaotic R\"ossler oscillators. Each elementary 
oscillator is described now by three variables ${\bf w} (t)=\{x(t),y(t),z(t)\}$,
and we have $f_x=-y-z,$ $f_y=x+ay$, and $f_z=b-cz+xz$, where $a$, $b$ and $c$ 
are fixed parameters. The collective dynamics of a globally coupled population
of $N$ identical R\"ossler oscillators ($i=1,2,...,N$) is governed by
equations 
\begin{equation}
\begin{array}{l}
\dot x_i=-y_i-z_i, \\ 
\dot y_i=x_i+ay_i+K(\bar y-y_i), \\ 
\dot z_i=b-cz_i+x_iz_i+K(\bar x\bar z-x_iz_i),
\end{array}
\label{1}
\end{equation}
where $\bar x(t)$, $\bar y(t)$ and $\bar z(t)$ are global averages, such
as $\bar x(t)=N^{-1}\sum_{i=1}^Nx_i(t)$, and $K$ is the intensity of
global coupling. This form of coupling is derived from (\ref{8}) by taking 
$\varepsilon =\varepsilon'=K$ and choosing a matrix $\hat{\bf A}$
whose elements are zero except for $A_{xy}=A_{xz}=-A_{yx}=1,$ $A_{zz}=c$,
and $A_{yy}=1-a$. The system (\ref{1}) reduces to a set of independent
identical R\"ossler oscillators \cite{Roess} for $K=0$.

The eigenvalues of the matrix $\hat{\bf A}$ are  $\lambda_{1}=c$ and 
$\lambda_{2,3}=\frac{1}{2}\left( 1-a\right) \pm \frac{1}{2}  
\sqrt{(1-a)^{2}-4}$. All three eigenvalues (or their real parts) are
positive if $a<1$. Therefore, for $K=1$ all deviations from the global
averages exponentially decrease with time and the states of all elements in
the population asymptotically converge. We see that at $K=1$ the oscillator
population (\ref{1}) possesses a global attractor which corresponds to
coherent motion of all elements of the system. This attractor coincides with
the attractor of a single R\"ossler oscillator and is chaotic for the
parameters $a=0.15$, $b=0.4$ and $c=8.5$ used in our simulations. Thus, the
population (\ref{1}) of globally coupled chaotic R\"ossler oscillators
should undergo condensation as the coupling intensity $K$ is gradually
increased towards $K=1$. To investigate the condensation phenomena, we have
performed numerical simulations of a population of $N=1000$ such oscillators
under varying intensity $K$ of global coupling. Below we first analyze  
statistical properties of the condensation transition and later consider how
condensation is influenced by heterogeneities.

To characterize condensation, distances 
$d_{ij}=[(x_i-x_j)^2+(y_i-y_j)^2+(z_i-z_j)^2]^{1/2}$
between the states of all $N(N-1)/2$ possible pairs of elements $(i,j)$ have
been computed. When condensation occurs, some of the elements have exactly
the same coordinates and distances between the states of these elements are
zero. The ratio of the number of pairs with zero distances to the total
number of pairs can thus be chosen as the {\it order parameter} $r$ of the
condensation transition. In absence of a condensate, $r=0$. On the other
hand, $r=1$ when complete mutual synchronization of the whole population
takes place.

To prepare the initial condition, the system was first allowed to evolve up
to $t=100$ without global coupling ($K=0$) so that its elements get
uniformly distributed over the R\"ossler attractor. Global coupling
was then introduced, time was reset to zero, and the evolution of the coupled
population was started. During a transient $T$, identical element pairs were
appearing. The transient was over when the number of these pairs did not
further increase with time. In our simulations, the transients were always
shorter than $T=1000$.

Figure 1 shows the dependence of the order parameter $r$ on the global
coupling intensity $K$ obtained by averaging over 20 realizations at $t=2000$. 
Condensation begins at $K_{c}\approx 0.017$ when a nonvanishing fraction
of identical pairs first appears. In the interval from $K_c$ to $K\approx  
0.06$, nonmonotonous dependence of the order parameter on the coupling
intensity is observed. At higher coupling intensities $K$, the fraction of
identical pairs steadily grows until full condensation ($r=1$) is
established for $K>K_0$, with $K_0\approx 0.093$. Under conditions of
partial condensation, the number of identical pairs varies from one
realization to another. Bars in Fig. 1 indicate statistical dispersion of
the simulation data at a few selected values of the global coupling
intensity. Note that, once a certain number of identical pairs has been
formed at the end of the transient period, it remains later constant in any
given simulation. Therefore, the dispersion reveals a degeneracy in the
asymptotic properties of this system in the partially condensed state.

The order parameter $r$ characterizes the overall degree of condensation,
but is not sensitive to the condensate structure. To determine the detailed
structure of the population, distributions over pair distances between the
states of all elements at a fixed time moment ($t=2000$) have been analyzed.
Histograms of such distributions at four different coupling intensities $K$
are shown in Fig. 2. They are constructed by counting pairs with distances
lying within subsequent intervals of width $\Delta d=0.25$. The number of
pairs inside each next interval, divided by the total number of pairs 
$N(N-1)/2$, yields the height of the respective bar. In absence of global
coupling, elements are relatively uniformly distributed over the
single-oscillator attractor and a smooth distribution over distances $d$ is
thus observed (Fig. 2a). When global coupling is introduced but remains
below the critical point $K_c$ of the condensation transition, the
distribution is modified (Fig. 2b). Now, pairs with small distances have
already appeared, though identical pairs are still absent (cf. Fig. 1).
Slightly above the critical point, a strongly nonuniform distribution with
many narrow peaks is seen (Fig. 2c). The distributions found at larger
intensities of global coupling (e.g., Fig. 2d) are formed by several
distinct lines. When complete condensation has taken place ($r=1$), the
histogram has only one line located at $d=0$ (not shown). Distributions
formed by several lines are characteristic for  situations when the whole
population breaks down into a number of coherent clusters. All elements
in a cluster follow the same dynamical trajectory and,
hence, distances between pairs of elements belonging to the same cluster are
zero. When elements from two different clusters are chosen, their
distances are all equal. If $M$ such clusters are present, the distribution
over pair distances would consist of $M(M-1)/2+1$ individual lines. Hence,
four lines seen in Fig. 2d correspond to a condensate with three
clusters. In our simulations, pure three-cluster regimes were typically
found starting from $K\approx 0.045$. They were replaced by pure two-cluster
regimes at $K\approx 0.058$. 

Hence, the condensate appearing above the transition point includes only a
small fraction of elements of the population, so that the order parameter is
small in this region. The condensed elements are distributed over a large 
number of small clusters. The rest of the population has asynchronous dynamics,
though it is already essentially influenced by global coupling between the
elements. When the coupling intensity is increased, the number of clusters
decreases. However, the typical sizes of clusters become at the same time
larger and therefore the condensate contains a greater fraction of the
population. Starting from a certain intensity of global coupling, all
elements belong to one of a few big clusters. As the coupling intensity is
further increased, relative sizes of clusters vary and some of them
disappear. This process eventually leads to complete condensation which can
be viewed as a single-cluster regime.

Fig. 3 shows the typical dependence of the order parameter $r$ of the
condensation transition on the total population size $N,$ where each point
is obtained by averaging over 20 independent realizations and intensity of
global coupling is kept constant ($K=0.08$). We see that a strong size
dependence is characteristic for relatively small populations ($N<500$).
Starting from a population size of about $N=1000$, the dependence displays
saturation, with the remaining small variations lying within the statistical
dispersion range. Hence, the results of our study of a population with 1000
oscillators might already be representative for the asymptotic behavior in
the infinite-size limit $N\to \infty$.

The complete condensation ($r=1$) has been found within a wide interval of
the global coupling intensity. Starting in this region with random initial
conditions for all elements, we see that the elements soon form a compact
cloud that rapidly shrinks with time. The characteristic radius $R(t)$ of
this cloud can be defined by 
\begin{equation}
R^{2}(t)={\frac{1}{N}}\sum_{i=1}^{N}\left[ \delta x_i^{2}(t)+\delta
y_i^{2}(t)+\delta z_i^{2}(t)\right] ,  \label{6}
\end{equation}
where $\delta x_i$, $\delta y_i$ and $\delta z_i$ are the deviations
from the respective global averages. The radius $R$ decreases with time, as
the trajectories of all elements asymptotically converge to the same orbit
(in our simulations the convergence was followed until the trajectories of
all elements became identical up to the computer precision of about $10^{-6}$). 
An important quantitative property of this regime is the characteristic
time which the system needs to reach the condensed state; the inverse of
this time represents the condensation exponent (closely related to the
transverse Lyapunov exponents discussed in Ref. \cite{Pec}): 
\begin{equation}
\gamma =-\lim_{t\rightarrow \infty }\frac{1}{t}\ln \frac{R(t)}{R(0)}
\end{equation}
The simulations show that $\gamma $ starts to grow from zero at $K=K_{0}$,
has a maximum at $K\approx 0.65$ and then slowly decreases approaching the
value $\gamma \approx 0.38$ at $K=1$.

The robustness of the complete condensation suggests that it might persist
in some form even in heterogeneous populations. To test this, we have
carried out simulations where the elements forming the population were not
identical. Heterogeneity was introduced by replacing the constant
parameter $c$ in Eqs. (\ref{1}) by random numbers $c_i$ that were
uniformly distributed inside the segment $\left[ c-\sigma_c,c+
\sigma_c\right] $. Alternatively, the same procedure was applied to the
parameter $a$. Starting from random initial conditions, a compact cloud of 
elements was again formed in the heterogeneous case.
However, the cloud did not shrink until it transformed into a single point.
Instead, after an initial decrese, its radius $R(t)$ fluctuated around a 
certain mean value. 
Fig. 4 shows the mean radius $\langle R\rangle$ as function of dispersions 
$\sigma_c$ or $\sigma_a$ for $K=0.2$. We see that in a wide interval of
heterogeneities the mean radius of the cloud is approximately linear
proportional to the dispersion. Moreover, even for relatively large
dispersions this radius is significantly smaller that the characteristic
attractor diameter $D$, representing the maximal possible pair distance at 
$K=0\ $(i.e. $D\approx 25$, cf. Fig. 2a). Therefore, even though the states
of all elements are no longer identical (i.e. the complete condensation is
destroyed), their variations remain small. The elements form a little cloud
whose center moves along a definite trajectory.

This behavior can be understood by analyzing the effect of weak
heterogeneities at the maximal strength of global coupling $K=1$. If the
parameter $c$ is replaced by $c_i=c+\Delta c_i$ with random variations 
$\Delta c_i$, the population dynamics is described by
\begin{equation}
\begin{array}{l}
\dot x_i=-y_i-z_i, \\ 
\dot y_i=x_i+(a-1)y_i+\bar y, \\ 
\dot z_i=b-c_iz_i+\bar x \bar z.
\end{array}
\label{inh}
\end{equation}
Hence, motion of each element is governed by linear equations of motion in
the presence of global driving forces $\bar y$ and $\bar x\bar z$ , which
are the same for all elements. Therefore, deviations $\delta z_i$ from the
global average obey the equation  
\begin{equation}
\dot{\delta z_i}=-c\delta z_i-\Delta c_i\bar z-\Delta c_i\delta
z_i+\overline{\Delta c_i\delta z_i}.
\end{equation}
If heterogeneities are weak, i.e. $\Delta c_i=\theta \zeta_i$ where 
$\theta$  is a small parameter, the last two terms in this equation have
order $\theta^2$ and can be neglected. Integrating the resulting linear
equation, we obtain 
\begin{equation}
\delta z_i(t)=\theta\zeta_i\int_0^\infty\exp (-c\tau )\bar z (t-\tau)\ d\tau .
\end{equation}
Considering the behavior of other deviations $\delta x_i$ and $\delta
y_i,$ one can similarly show that they are also proportional to $\Delta 
c_i=\theta \zeta_i$. This means that the radius $\langle R\rangle $ of
the population cloud depends as $\langle R\rangle \sim \theta $ on the
heterogeneity strength. Note also that in this case all deviations are
scaling-identical, i.e., for example, $\delta z_i(t)=\theta \zeta_iQ(t)$
where the function $Q(t)$ is the same for any element $i$. The same results
are obtained when weak random variations of the parameter $a$ are considered.

Our simulations have revealed that the linear dependence of the mean radius 
$\langle R\rangle $ on the dispersions $\sigma_c$ and $\sigma_a$, that
has been analytically shown above only at $K=1$, persists in a wide range of 
coupling intensities. Moreover, this dependence is found even for
relatively large dispersions, where the parameter $\theta$ would not be
small. This behavior is apparently a consequence of a strong compression
provided in the entire interval of complete condensation by global coupling.

The condensation transition has been studied above for a particular kind of
chaotic oscillators and for a certain form of global coupling. We have
however checked that similar results are also obtained when a different
(`vector') form of global coupling is used. This suggests that observed
statistical properties may be typical for various large populations of
globally coupled chaotic oscillators undergoing a condensation transition.

Financial support from Fundaci\'on Antorchas, Argentina, and from
Alexander von Humboldt-Stiftung, Germany, is gratefully acknowledged. D.H.Z.
wishes to thank the Fritz Haber Institute for hospitality during his stay in
Berlin.

\newpage

\begin{figure} 
\caption{Order parameter $r$ of the condensation transition as function of
the coupling constant $K$. Averaging over 20 independent realizations is
performed;   bars indicate   statistical dispersion of data at several
selected points.}
\end{figure}

\begin{figure} 
\caption{Normalized histograms of the distribution over pair distances $d$
between elements in a population of $N=1000$ globally coupled R\"ossler
oscillators at different stages of the condensation transition: (a) $K=0$;
(b) $K=0.01$; (c) $K=0.02$; (d) $K=0.05$.}
\end{figure}

\begin{figure} 
\caption{Dependence of the order parameter $r$ on the population size $N$ at
a fixed intensity of global coupling $K=0.08$. Averaging over 20 independent
realizations;   bars indicate   statistical dispersion of data. }
\end{figure}

\begin{figure} 
\caption{Radius $\langle R\rangle$ of the cloud formed by a heterogeneous 
population as function of dispersions $\sigma_a$ (empty circles) and $\sigma_c$
(filled circles) at $K=0.2$.}
\end{figure}

\end{document}